# Molecular contrast on phase-contrast microscope

Keiichiro Toda,[1,†] Miu Tamamitsu,[1,†] Ryoichi Horisaki,[2,3] and Takuro Ideguchi[1,3,*]

[1]Department of Physics, The University of Tokyo, Tokyo 113-0033, Japan
[2]Graduate School of Information Science and Technology, Osaka University, Osaka 565-0871, Japan
[3]PRESTPO, Japan Science and Technology Agency, Saitama 332-0012, Japan
[*]Corresponding author: ideguchi@phys.s.u-tokyo.ac.jp
[†]These authors contributed equally to the work.

**Abstract**
An optical microscope enables image-based findings and diagnosis on microscopic targets, which is indispensable in many scientific, industrial and medical settings. Majority of microscope users are accustomed to standard benchtop microscopes, but it fails to give molecular contrast of the specimen which otherwise requires expensive specialized instruments to measure, accompanied by chemical or optical damages to the sample and/or slow imaging speed. Here, we report on a simple optical instrument, comprising of a semiconductor amplitude-modulated mid-infrared quantum cascade laser, that is attached to a standard microscope to deliver the additional molecular contrast of the specimen on top of its conventional microscopic image. We attach this instrument, termed molecular-contrast unit, to a standard phase-contrast microscope, and demonstrate high-speed label-free molecular-contrast phase-contrast imaging of biological cells under low-power optical illumination. Our simple molecular-contrast unit can empower existing standard microscopes of various users, by delivering a convenient accessibility to the molecular world.

An optical microscope is a universal tool permeated through numerous aspects of science, industry and medicine. Observation of microscopic world is at the basis of studying unknown physical [1-3] and biological [4-6] phenomena, daily research activities, mass-production inspection such as semiconductor [7] and pharmacy [8] production lines, and clinical and medical diagnosis [9-11], to just name a few. Majority of microscope users are not an expert in microscopic technologies, and are accustomed to standard benchtop platforms. Today, various microscopic modalities are offered by commercial benchtop platforms including bright-field, dark-field, differential-interference-contrast, and phase-contrast (PC) [12] modalities. These are powerful tools to obtain image-contrast based on optical absorption, scattering or thickness revealing the microscopic physiology of the specimen, and yet convenient because the specimen can be observed, without any specific preparation or alteration, in real time or even at a high frame rate if equipped with a commercial high-speed camera.

However, when one's interest comes to molecular contrast (MC), the application-variety and/or ease-of-use of an optical microscope becomes strictly limited. Fluorescence microscopy [5], for example, is a wide-spread molecular imaging technique mostly used in the field of biology, but requires chemical alteration and/or photodamage [13] to the specimen by introducing an exogenous labelling agent called fluorescent probe. In many applications, including biology, such alteration of the sample is not permitted or demanded. Label-free alternatives would be molecular-vibrational microscopy, finding applications in biology [14-18], medicine [19-21], pharmacy [22-24], planetary science [25], archaeology [26], safety inspection [27], etc., where optical absorption or scattering caused by sample-intrinsic molecular vibrations is measured. These methods, however, suffer from low-spatial resolution in the case of mid-infrared (MIR) absorption imaging [28], and low-sensitivity in the case of Raman-scattering imaging resulting in slow imaging speed and/or high photodamage to the sample [29]. Equally important is that these systems typically require expensive, large-scale, custom-made optical systems, which is one of the major factors that interferes with technological transition to non-experts and hence prevents its wide-spread use to various other fields.

Here, we propose and demonstrate a concept of adding MC to a benchtop optical microscope system, by attaching an add-on "MC unit" to an existing microscope platform comprising of a frequency-tunable semiconductor amplitude-modulated MIR quantum cascade laser (QCL) [30]. The amplitude-modulated MIR laser beam illuminated onto the entire specimen region generates the refractive-index modulation via photothermal effect [17,18,24] at specific sites where vibrationally resonant molecules exist, which is then visualized by phase-sensitive imaging with a conventional PC microscope. Our system, which we call MC-PC microscope, offers (1) simple and easy implementation that allows non-experts to have MC on their conventional microscope images by simply attaching an MC unit to their existing PC microscopes, and yet (2) high performance in terms of a spatial resolution based on the diffraction limit of the visible light used for standard PC microscopy while at the same time a high molecular sensitivity based on MIR absorption (rather than Raman scattering) and phase (rather than, e.g., amplitude) detection, allowing for high-speed molecular spectroscopic imaging under low-power optical illumination reducing photodamage to the sample.

Our MC-PC microscope is realized by the MC unit, comprising of an amplitude-modulated MIR QCL, attached to a standard PC microscope platform (see Fig. 1). The amplitude-modulated continuous-wave MIR beam excites the whole area of the sample at the objective focus. The molecular-vibrational photothermal effect takes place only at specific sites in the sample where vibrationally resonant molecules exist, which appears as temporal intensity modulations in the PC image. The camera records the video of this phenomenon, and the resulting periodic photothermal signals are computationally extracted to produce the MC image. Note that a PC microscope, unlike, e.g., a bright-field microscope, is the key instrument to achieve the sensitive detection of the MC, as the photothermal refractive-index change appears in the optical phase rather than in the amplitude.

To demonstrate the high frame rate and bond-specific molecular contrast of our MC-PC microscope, we measure a mixture of polystyrene and porous silica beads of size ~ 5 μm immersed in index-matching oil with the MIR beam configured to 1,045 cm$^{-1}$ (resonant to Si-O-Si stretch of $SiO_2$). Figure 2a shows the temporal PC intensities of a polystyrene and a porous silica bead. The former stays constant while the latter shows periodic modulation at a frequency of 3,000 Hz which is the modulation frequency of the MIR beam. This clearly shows the modulated MIR beam induces the bond-specific PC intensity modulation. Shown in Fig. 2b is the MC image obtained in one MIR excitation cycle, visualizing the two-dimensional locations of the porous silica beads. The wide-field image-acquisition speed is 3,000 fps over ~ 200 × 200 of diffraction-limited pixels (~ 100 μm × 100 μm area), which is orders of magnitude higher than those of other high-speed molecular-vibrational imaging techniques such as coherent Raman imaging [16] and could be further increased with a higher MIR modulation frequency and higher-power MIR light sources (see "MC frame rate and SNR" section in Supplementary Information). Finally, Fig. 2d shows the MC-PC image synthesized by overlaying the obtained MC on top of the conventional PC image, where the porous silica beads are selectively highlighted.

Now we show the ability of our MC-PC microscope to perform molecular-vibrational spectroscopic imaging by obtaining the MC images of Henrietta Lacks (HeLa) cells at various MIR wavenumbers ranging from 1,492.5 to 1,615 cm$^{-1}$. Figure 3a shows normalized MC spectra measured at three different spatial points in the cells indicated by the red, green and blue arrows in Fig. 3b. Note that the MC values are normalized by the corresponding MIR power (see Fig. S1 for the linearity of MC with MIR power). The blue curve represents a characteristic spectrum of a HeLa cell [31], where the absorption peak found at 1,530 cm$^{-1}$ represents the amide II band and the absorbance-increase from 1,605 to 1,615 cm$^{-1}$ that of amide I band of peptide bonds characteristic to proteins. On the other hand, the red and green curves show peculiar peaks at 1,500 and 1,554 cm$^{-1}$, respectively. Although we have not assigned the origins of these spectral features due to the lack of literature, we could have a chance to find new information.

Our MC-PC microscope offers the spatial resolution of visible light (in this case ~1 μm, determined by the objective NA 0.6 and wavelength 630 nm) which is an order of magnitude higher than that of MIR absorption imaging (~ 10 μm). A higher resolution can be easily achieved with a higher objective NA and a shorter wavelength, e.g., 530 nm with a NA 0.85 and a wavelength 450 nm. The high spatial resolution allows us to detect small (i.e., few micrometers) spectral variations in the MC spatial distribution. Figure 3c represents MC images of the HeLa cells measured at 1,500, 1,530, 1,554 and 1,615 cm$^{-1}$. The MC images at 1,530 and 1,615 cm$^{-1}$ show nearly identical profiles, indicating these MCs originate in the same type of molecular species (possibly, proteins in this case), while the other two images at 1,500 and 1,554 cm$^{-1}$ show characteristic distributions, respectively. Shown in Fig. 3d is an overlay of the three MC images at 1,500, 1,530 and 1,554 cm$^{-1}$ on top of the standard PC image, highlighting the different MC spatial distributions.

In terms of the sensitivity and acquisition-speed, each MC image is obtained at 10 fps (100 ms acquisition time) with the highest signal-to-noise ratio (SNR) of 12.6. Since our microscope is based on a standard PC microscope, the energy of the visible probe light, which is well known to induce photodamage to biological samples [13,29], is orders of magnitude lower (~ 630 pJ/μm$^2$) than those of other molecular imaging techniques such as fluorescence and Raman imaging. The frame rate can be further increased at the expense of the SNR.

It is of great importance that the dual-modal MC-PC imaging capability is a unique feature of our microscope which allows us to determine the molecular distributions within the global physiology of the sample from a single dataset. This is not readily achieved with conventional imaging techniques as separate imaging modalities are traditionally used for optical-phase-sensitive (i.e., physiological) [4] and molecular-sensitive imaging [5,16]. In Fig. 3d, for example, the protein concentration (blue) can be observed to be higher at the center and around the nucleus of the cells while distributions of the other molecular species seem to follow some cellular structures recognized in the PC image (see, e.g., the green and red contrasts found on the protruding structures recognized at the bottom right corner of the top cell and the left side of the bottom cell, respectively). Such physio-chemical analysis would be more interesting with distinct biological contrasts such as MCs resonant to lipids and deoxyribonucleic acids (DNAs), which is readily achievable by implementing MIR light sources of other wavenumber ranges (e.g., ~ 1,740 cm$^{-1}$ for >C=O ester stretch representing lipids, ~ 1,080 or 1,240 cm$^{-1}$ for O-P=O stretch representing DNAs, etc.) [32]. Overall, this biological demonstration verifies the ability of our MC-PC microscope to perform high-speed, high-resolution and low-power vibrational-molecular spectroscopic imaging.

Several technical improvements could be considered to make our system more robust and convenient. It is desirable to homogenize the spatial profile of the MIR beam at the sample plane so that the systematic distribution of the MC is physically removed, which could be realized using, e.g., diffractive-optical elements [33]. Another possible improvement is to collinearly introduce the MIR excitation and the visible probe beams using, e.g., a reflective condenser lens, for easier and possibly automated alignment of the MIR beam. Our MC unit could be packed with a reflective condenser lens and a probe light source to serve as a compact add-on module to existing optical microscope platforms. The sensitivity of our MC microscope could also be further enhanced. Standard benchtop optical microscopes are typically shot-noise limited due to the active illumination resulting in a large number of detectable photons, suggesting that the camera's full-well capacity and frame rate become the ultimate limitations. Technological advancement in the camera industry would then lead to an enhanced sensitivity with the MC microscopes. We also expect other imaging modalities could be used to enhance the spatial resolution and/or achieve quantitative measurement of optical phase, such as dark-field, differential-interference-contrast and other quantitative phase imaging techniques, some of which can also be operated on a standard benchtop optical microscope [4].

## Methods
### Principle of the MC-PC microscope
We used the following instruments to construct our MC-PC microscope: (1) PC microscope: LUCPLFLN 40XPH and IX73 (Olympus), (2) MIR QCL light sources: QD9500CM1 (Thorlabs) configured to 1,045 cm$^{-1}$ for experiments with bead samples and DO418, Hedgehog (Daylight Solutions) for spectroscopic experiments with HeLa cells (spectral coverage: 1,450 – 1,640 cm$^{-1}$, spectral resolution: ~ 1 cm$^{-1}$, spectral tuning speed: > 1,000 cm$^{-1}$/s), (3) camera: MEMRECAM HX-7s (Nac) and (4) visible LED light source: SOLIS-623C (Thorlabs). The MIR beam is weakly focused onto the sample placed at the objective focus using a CaF$_2$ lens. We avoid tight focusing of the MIR beam in order to illuminate and excite the entire region of the sample (~100 μm × 100 μm). The sample is sandwiched between two 500-μm-thick CaF$_2$ substrates in order to avoid absorption of MIR light by the substrate medium. The amplitude of the MIR light is modulated either electrically (QD9500CM1) or using an optical chopper (DO418) at a desired rate.

### Experimental conditions
We used the following materials for our experiments: (1) 4.8-μm polystyrene beads: 17135-5, Polybead Microspheres (Polysciences, Inc.), (2) 5-μm porous silica beads: 43-00-503, Sicastar (micromod Partikeltechnologies GmbH), and (3) index-matching oil: refractive index 1.50 at 587.56nm (SHIMAZU). The visible LED power is ~ 50 μW at the sample plane.

### MC image synthesis procedure
The MC image is obtained from the time-sequence of MIR-modulated PC images by calculating the difference of two averaged images; the ones calculated from the time frames showing the lower-half (i.e., MIR-OFF image) and the higher-half (i.e., MIR-ON image) magnitudes of the photothermal PC modulation at the MIR absorbers. The spurious negative signal in the MC image (which does not directly originate from the photothermal effect) is eliminated using the negative-contrast filter described in Supplementary Information (see Fig. S3).

### High-speed imaging on the mixture of beads
The MIR modulation frequency is 3,000 Hz and the camera frame rate is 10,000 fps. To derive the SNR, the noise level is determined by the standard deviation of the empty region (20 × 20 pixels, 12 μm × 12 μm) of the MC image before the negative-contrast filtering, while the signal level is determined by the averaged MC at the center of the beads (5 × 5 pixels, 3 μm × 3 μm) where the strongest MC is observed. The MIR power is ~ 10 mW.

### Spectroscopic imaging of HeLa cells
The HeLa cells are immersed in deuterium oxide. The MIR modulation frequency is 250 Hz and the camera frame rate is 10,000 fps. A continuous series of 1,000 images is used to calculate each MC image, corresponding to 25 cycles of MIR modulation and the acquisition time of 100 ms. The SNR is calculated based on the MC image at 1,530 cm$^{-1}$. The noise level is determined by the standard deviation of the empty region (~ 25 × 25 pixels, 14 μm ×14 μm) in the MC image before the negative-contrast filtering, while the signal level is determined by the averaged MC at the center of the HeLa cell (~ 5 × 5 pixels, 3 μm × 3 μm) where the strongest MC is observed. The spectral resolution is determined by the wavelength-tunability of the light source, which is ~ 1 cm$^{-1}$ in our case although we use 10 cm$^{-1}$ increment in this experiment. The MIR power is dependent on the wavenumber and varies between ~ 16 - 40 mW.

## Acknowledgements
We thank Makoto Kuwata-Gonokami and Junji Yumoto for letting us use their equipment. We are grateful to Yu Nagashima for offering biological cells. This work was financially supported by JST PRESTO (JPMJPR17G2), JSPS KAKENHI (17H04852, 17K19071), Research Foundation for Opto-Science and Technology, and The Murata Science Foundation.

## Author contributions
T.I. conceived the work. K.T., M.T., R. H. and T.I. designed the system. K.T. and M.T. conducted the experiment and analyzed the data. T.I. supervised the work. M.T. and T.I. wrote the manuscript and all authors contributed to improvement of the manuscript.

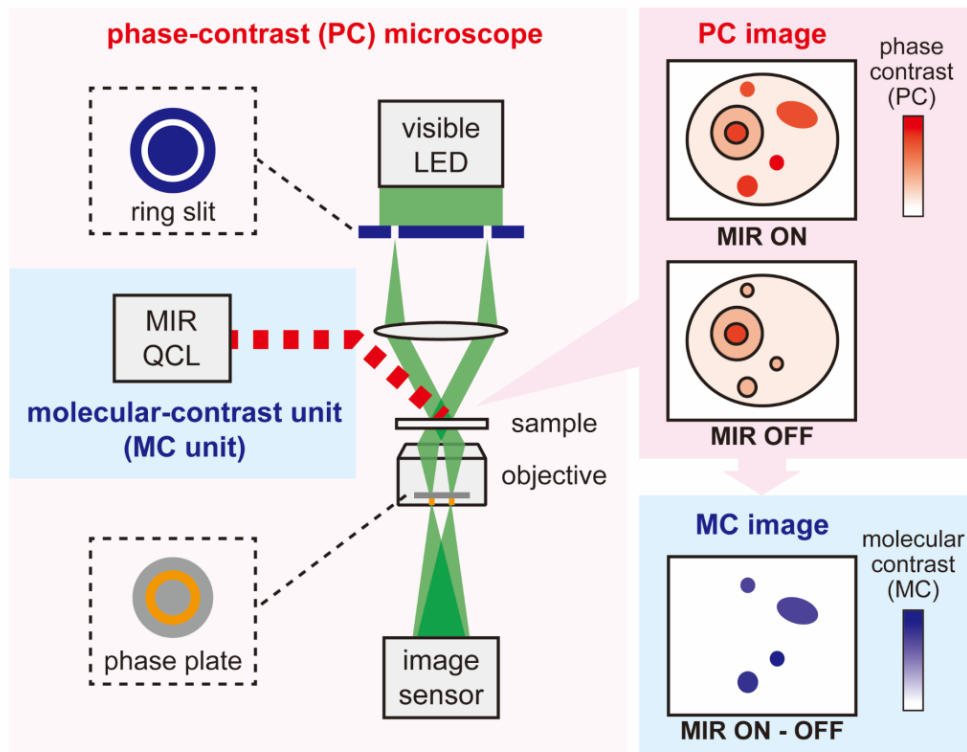

**Figure 1 | Molecular-contrast phase-contrast (MC-PC) microscopy.** To obtain the MC image of the sample, the add-on "MC unit" is attached to an existing standard PC microscope platform, which consists of an amplitude-modulated continuous-wave MIR QCL. The amplitude-modulated MIR beam is weakly focused onto the sample placed at the objective focus to cover the entire region of the specimen. This induces the molecular-vibrational photothermal effect at specific sites within the specimen where vibrationally resonant molecules exist, which is detected as temporal intensity-modulations in the time-series of the recorded PC images. These photothermal signatures are computationally extracted to reveal and add the resonant MC to conventional PC microscope images.

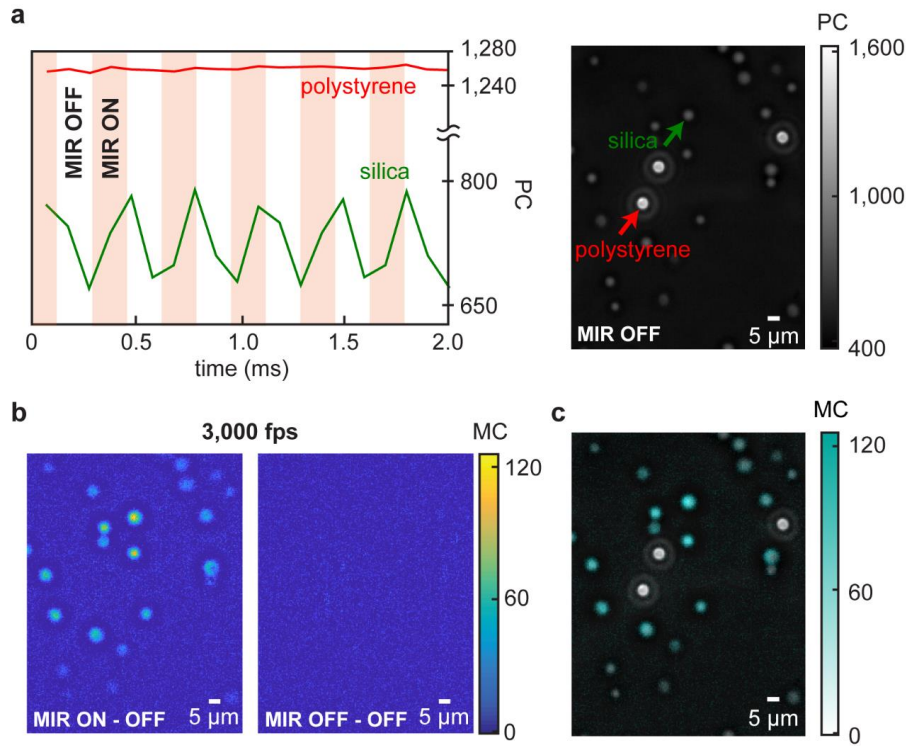

**Figure 2 | High-speed wide-field bond-specific imaging with the MC-PC microscope. a**, Site-specific time-dependent PC signals. The shown signals are the spatial-average of 6 × 6 pixels (3 μm × 3 μm) at the center of the silica (green curve and arrow) and the polystyrene (red curve and arrow) beads. The right panel shows a standard PC microscope image. **b**, MC image obtained at 3,000 fps, corresponding to one MIR excitation cycle. The SNR is 13.2. Note that there exists a systematic spatial variation of the MC due to the spatial intensity profile of the excitation MIR beam spot at the sample plane. Left: MIR ON – OFF image. Right: MIR OFF – OFF image. **c**, MC-PC image synthesized by overlaying the MC (obtained in b) on top of the conventional PC image (shown in a). The MC selectively highlights the locations of the resonant porous silica beads out of the mixture sample, providing additional but powerful information on top of the conventional PC image.

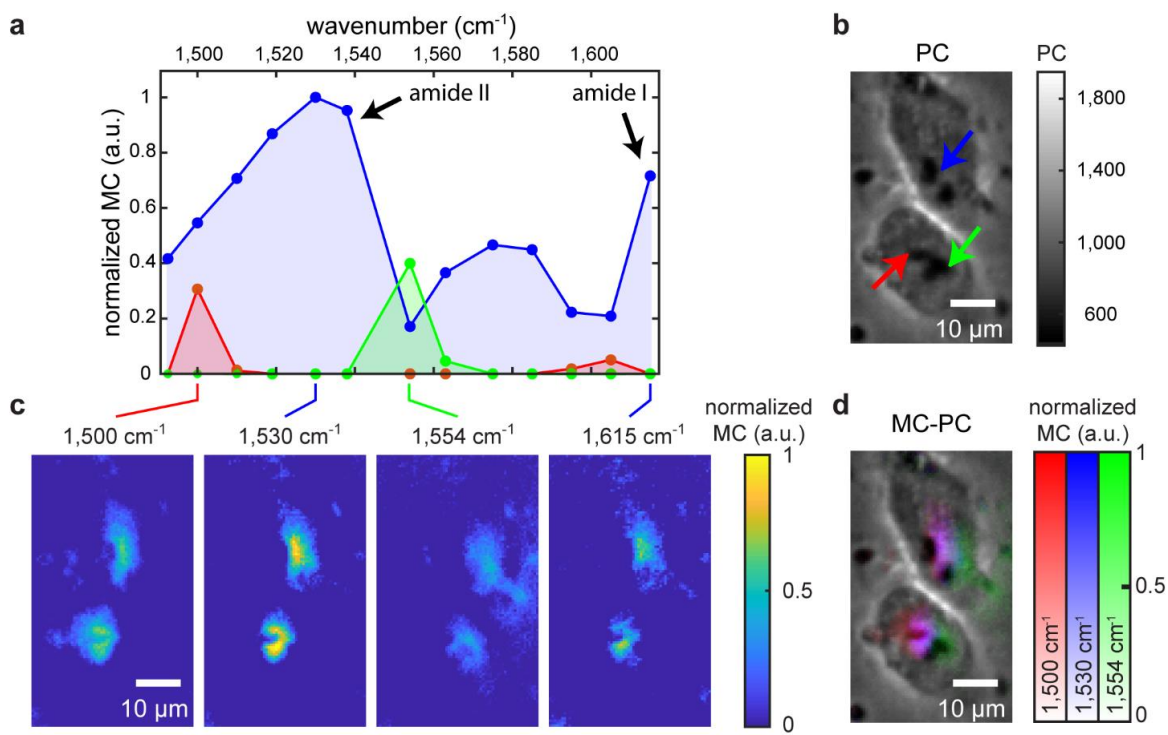

**Figure 3 | Molecular-vibrational spectroscopic imaging of HeLa cells measured with the MC-PC microscope. a**, MC spectra of HeLa cells obtained at the spatial points indicated by the arrows in **b**. The colors of the data points correspond to those of the arrows in **b**. **b**, PC image of the HeLa cells obtained at the MIR-OFF state, similar to a standard PC microscopic image. **c**, MC images of the HeLa cells measured under the vibrational excitation by the MIR beam lasing at 1,500, 1,530, 1,554 or 1,615 $cm^{-1}$. Each MC image is obtained at 10 fps (i.e., 100 ms acquisition time). **d**, MC-PC image of the HeLa cells comprised of the MC images at 1,500, 1,530 and 1,554 $cm^{-1}$.

## Supplementary Information

## Linearity of MC vs MIR excitation power

To verify the response of the MC with respect to the MIR excitation power, we obtain the MC images of HeLa cells immersed in deuterium oxide at 1,554 cm$^{-1}$ with various MIR powers ranging between ~ 5 – 27 mW over the area of ~ 100 μm × 100 μm, and plot the obtained MC. The MIR modulation frequency is 250 Hz and the camera frame rate is 10,000 fps. A continuous series of 10,000 images is used to calculate each MC image, corresponding to 250 cycles of MIR modulation and the acquisition time of 1 s. The MC in each image is calculated by averaging 2 × 2 pixels (~ 1 μm × 1 μm) region of the center part of the cell. The result shown in Fig. S1 verifies the linear relation.

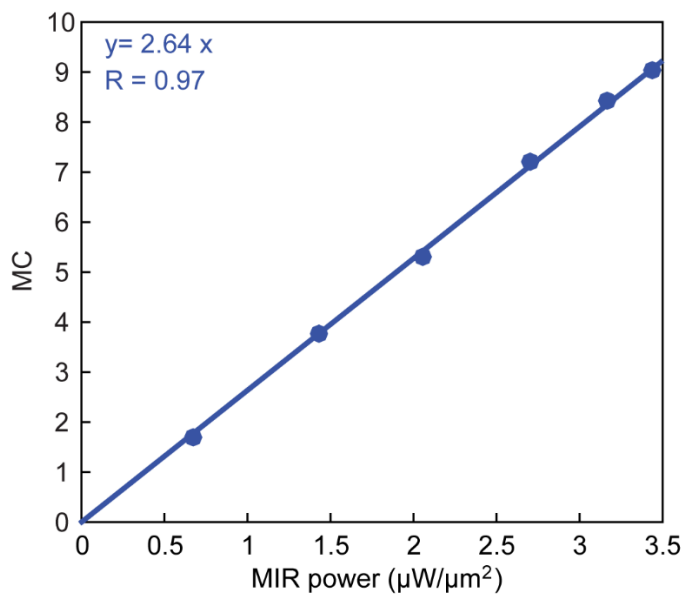

**Figure S1 | Linearity of MC vs MIR excitation power with the MC-PC microscope.**

## MIR modulation frequency vs MC, frame rate and spatial resolution

We characterize the performance of our MC-PC microscope in terms of the MC, frame rate and spatial resolution. Shown in Fig. S2 is how the MC and the full-width-at-half-maximum (FWHM) width of a 5 μm porous silica bead in the index-matching oil change with different MIR modulation frequencies (or MIR exposure time per modulation cycle). Note that the modulation frequency translates to the maximum possible frame rate to obtain the MC image. For each MIR modulation frequency, the bead's MC image is retrieved from a continuous series of 5,000 PC images captured at the frame rate of 100,000 fps (measurement time 0.05 s). This gives the same noise floor for the MC image of each MIR modulation frequency. To retrieve the peak contrast and the FWHM width, each of the obtained MC images is interpolated by a factor of 4 in each spatial dimension, to create its smooth cross-sectional curve. If the frame rate is kept the same, the maximum possible frame rate is in trade off with the SNR, e.g., when capturing at 100,000 fps, 1,000 Hz MIR modulation frequency translates to 1,000 fps at maximum for the MC-image acquisition by averaging 100 frames, whereas 100 Hz MIR modulation frequency translates to 100 fps at maximum for the MC-image acquisition by averaging 1,000 frames.

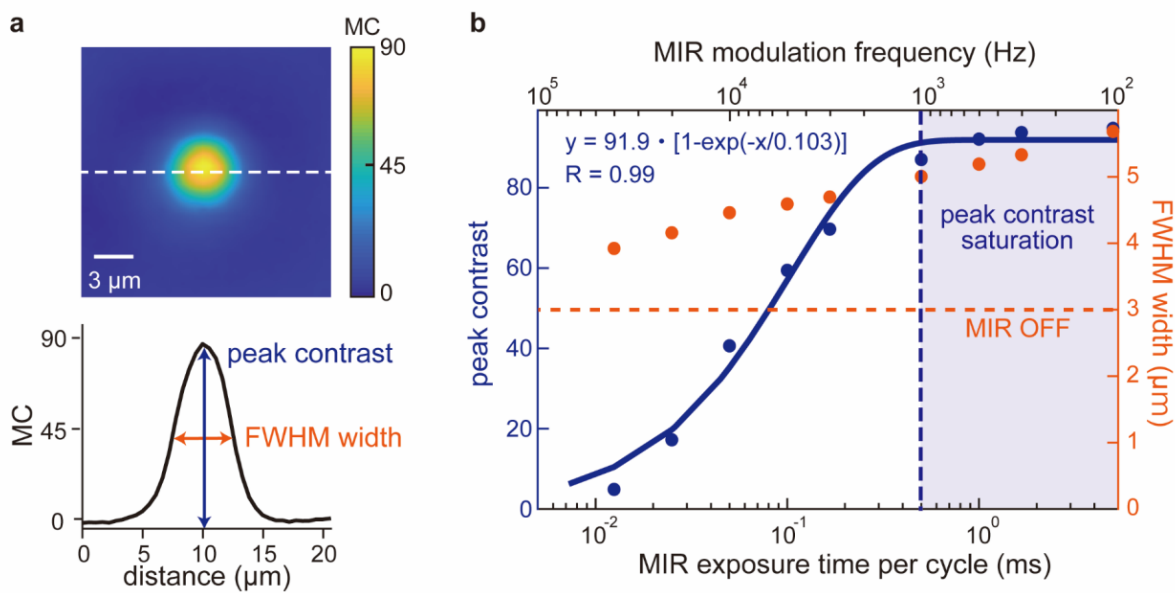

**Figure S2 | MIR modulation frequency vs MC, frame rate and spatial resolution. a**, MC image of a 5-μm porous silica bead immersed in the index-matching oil used to characterize the system performance. The peak contrast and the FWHM width are used to analyze the system performance. **b**, Performance characterization of the MC-PC microscope. Blue dots: peak contrast. Orange dots: FWHM width of MC images. Orange dashed line: FWHM width of the standard PC image at the complete MIR-OFF state. Blue dashed line: MIR frequency at which the peak-contrast saturates.

The MC increases with a longer MIR exposure and saturates above 0.5 ms, due to the existence of the heat diffusion time $\tau \propto \frac{1}{\alpha}$ ($\alpha$: thermal diffusivity of the heat absorber) [S1,S2]. When the MIR exposure is longer than $\tau$, the supplied heat accumulates to balance with the heat diffusion, building a strong saturated MC. Such phenomenon can be modeled by a rate equation, and our measurement data indeed fits well to an exponential function ($\tau \sim 100$ μs), suggesting that ~ 1,000 Hz would be the optimum MIR modulation frequency in terms of the MC and frame rate for an object ~ 5 μm, which could vary depending on the size of interest. Many liquid, polymer and glass materials have $\alpha \sim 10^{-7}$ [m$^2$/s] (e.g., paraffin: ~ 0.8 x 10$^{-7}$, ethanol: ~ 0.85 x 10$^{-7}$, water: ~ 1.4 x 10$^{-7}$, polycarbonate: ~ 1.5 x 10$^{-7}$, glass: ~ 5.6 x10$^{-7}$) [S3-S5], suggesting that similar results could be replicated in other experimental conditions. On the other hand, the MC FWHM width becomes larger than that of the original PC image (~ 3 μm) and increases with a longer MIR exposure, due to the heat diffusion towards the surrounding medium. Notice, however, that the standard PC images obtained at the MIR-OFF-state can serve as the high-resolution morphological support on which the MC can be overlaid to highlight the additional molecular information.

**Filtering spurious negative signal in MC image**
In our MC-PC measurement, a spurious negative signal is observed in the MC image. This negative MC is not the direct consequence of the photothermal effect but other phenomena such as Halo effect accompanied by the PC measurement. Figure S3a and b show the PC and MC images of the HeLa cells discussed in Fig. 3 in the main text, respectively. As seen in the MC image, we can clearly see the negative-MC region around the positive-MC region. To visualize the direct photothermal effect (the positive MC) only, we filter off the spurious negative contrast with a binary mask shown in Fig. S3c. The filtered MC is shown in Fig. 3c in the main text.

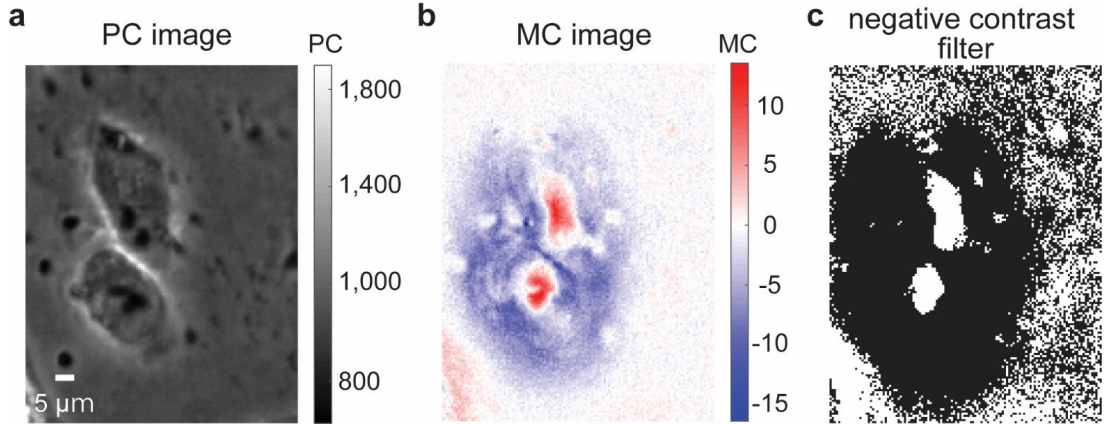

**Figure S3 | Filtering spurious negative signal in MC image. a**, Standard PC microscopic image of the HeLa cells. **b**, MC image. **c**, Negative-contrast filter (binary mask).

**MC frame rate and SNR**
The limitation of the MC frame rate, $F_{MC}$, is given by the camera frame rate, $F_{cameara}$, and the MIR modulation frequency, $F_{MIR}$, by the following equation:

$$F_{MC} \leq F_{MIR} \leq F_{cameara}/2. \tag{S1}$$

The first inequality represents that at least one cycle of MIR modulation is necessary to obtain the MC image, whereas the second the Nyquist-Shannon theorem for sampling the MIR modulation with the camera. Therefore, the ultimate limitation of the MC frame rate is determined by the camera frame rate. The higher MC frame rate, however, generally results in a lower SNR. In terms of the noise, the noise level increases with an increased MC frame rate, due to reduction in the number of averaged frames per MIR excitation cycle. In terms of the signal, the MC shows exponential decrease with a higher MIR modulation frequency below the saturation level as shown in Fig. S2. To increase the SNR, the following modifications can be made. For a fixed MIR modulation frequency, a higher MC can be obtained with a higher MIR power (realized by e.g., having a higher output from the MIR light source, focusing the MIR beam tightly, etc.) as the photothermal optical-phase change is linear to the input MIR power (see Fig. S1). The higher SNR can also be obtained with image sensors with a higher full-well capacity in the condition dominated by the optical shot noise.

**References**
S1. Zharov, V. P., Mercer, K. E., Galitovskaya, E. N., & Smeltzer, M. S., "Photothermal nanotherapeutics and nanodiagnostics for selective killing of bacteria targeted with gold nanoparticles," *Biophysical journal* **90**, 619 (2006).
S2. Andika, M., Chen, G. C. K., & Vasudevan, S., "Excitation temporal pulse shape and probe beam size effect on pulsed photothermal lens of single particle," *JOSA B* **27**, 796 (2010).
S3. Blumm, J., & Lindemann, A., "Characterization of the thermophysical properties of molten polymers and liquids using the flash technique," *High Temp. High Press* **35**, 627 (2003).